\documentclass[aps,prb,groupedaddress,showpacs,twocolumn,superscriptaddress]{revtex4}
\usepackage{graphicx}
\usepackage{amsmath}
\usepackage{amssymb}
\usepackage{bbm}
\usepackage{xspace}
\usepackage{color}
\usepackage{footnote}
\usepackage{comment}
\usepackage{placeins}%enables \FloatBarier
\usepackage{hyperref}

\newcommand{\phys}{} %my preference

\definecolor{orange}{RGB}{252,77,6}
\definecolor{brown}{RGB}{200,127,50}

\begin{document}

\title{Impact ionization processes  in the steady state of a driven Mott insulating layer coupled to metallic leads}

\author{Max E. Sorantin}
\email[]{sorantin@tugraz.at}
\affiliation{Institute of Theoretical and Computational Physics, Graz University of Technology, 8010 Graz, Austria}
\author{Antonius Dorda}
\affiliation{Institute of Theoretical and Computational Physics, Graz University of Technology, 8010 Graz, Austria}
\author{Karsten Held}
\affiliation{Institute of Solid State Physics, TU Wien, 1040 Vienna, Austria}
\author{Enrico Arrigoni}
\email[]{arrigoni@tugraz.at}
\affiliation{Institute of Theoretical and Computational Physics, Graz University of Technology, 8010 Graz, Austria}

\date{\today}

\begin{abstract}

We study a simple model of photovoltaic energy harvesting across a Mott insulating gap consisting of a correlated layer connected to two metallic leads held at different chemical potentials. We address in particular the issue of impact ionization, whereby a particle photoexcited to the high-energy part of the upper Hubbard band uses its extra energy to produce a second particle-hole excitation. We find a drastic increase of the photocurrent upon entering the frequency regime where  impact ionization is possible.
At large values of the Mott gap, where impact ionization is energetically not allowed, we observe a suppression of the current and  a piling up of charge in the high-energy part of the upper Hubbard band.Our study is based on a Floquet dynamical mean field theory treatment of the steady state with the so-called auxiliary master equation approach as impurity solver. We verify that an additional approximation, taking the self-energy diagonal in the Floquet indices, is appropriate 
for the parameter range we are considering.

\end{abstract}

%71.10.-w Theories and models of many-electron systems
%71.15.-m Methods of electronic structure calculations
%71.27+a Strongly correlated electron systems; heavy fermions
%72.15.Qm Scattering mechanisms and Kondo effect
%73.21.La 	Quantum dots (Electron states and collective excitations in multilayers, quantum wells, mesoscopic, and nanoscale systems)
%73.23.-b       Electronic transport in mesoscopic systems
%73.63.Kv       Quantum dots (Electronic transport in nanoscale materials and structures)
\pacs{71.15.-m,71.27+a,72.15.Qm,73.21.La,73.63.Kv}

\maketitle
\section{Introduction}\label{sec:introduction}
Strongly correlated materials are known to display intriguing effects and show properties not observed in ordinary systems.
Some examples are high-temperature superconductivity~\cite{ke.ki.15}, half-metallicity~\cite{co.ve.02}, spin-charge separation~\cite{ki.ma.96} and the Kondo effect~\cite{hews} 
just to quote a few. A prototypical class of these materials are so called Mott insulators where strong electronic interactions are responsible for the spectral gap as realized in transition metal oxides (TMOs). Recent theoretical works have proposed these materials as candidates for  efficient photovoltaics~\cite{mano.10,co.ma.14,as.bl.13}, exploiting electronic correlations  to increase the photovoltaic efficiency. 

The key idea is that in a strongly correlated insulator high energy electrons, created by photo excitation, are likely to undergo a process called impact ionization thereby exciting another electron across the gap. Although impact ionization is also present in conventional semiconductor devices, the time scales are such that a highly excited electron will generally dissipate its energy to phonons. In contrast,  the time scale of electron-electron (e-e) scattering is orders of magnitude shorter in correlated TMOs because of the strong interaction. In this way, the  excess energy of photo-excited electrons is substantially less prone to thermal losses and the efficiency of the resulting solar cell is not restricted by the Shockley-Queisser limit~\cite{sh.qu.61} any longer. Previous works have studied Mott systems after a photoexcitation with time-dependent dynamical mean field theory (t-DMFT) investigating the role of impact ionization~\cite{ec.we.11,we.he.14} as well
 as doublon dynamics~\cite{ec.we.13,ec.we.14} in the subsequent thermalization. This work confirmed the dominant character of impact ionization on short timescales of the order of ten femto seconds~\cite{we.he.14} and a high mobility of charge carriers in layered structures~\cite{ec.we.14}.

\begin{figure}[h]
\begin{center}
 \includegraphics[width=0.75\columnwidth]{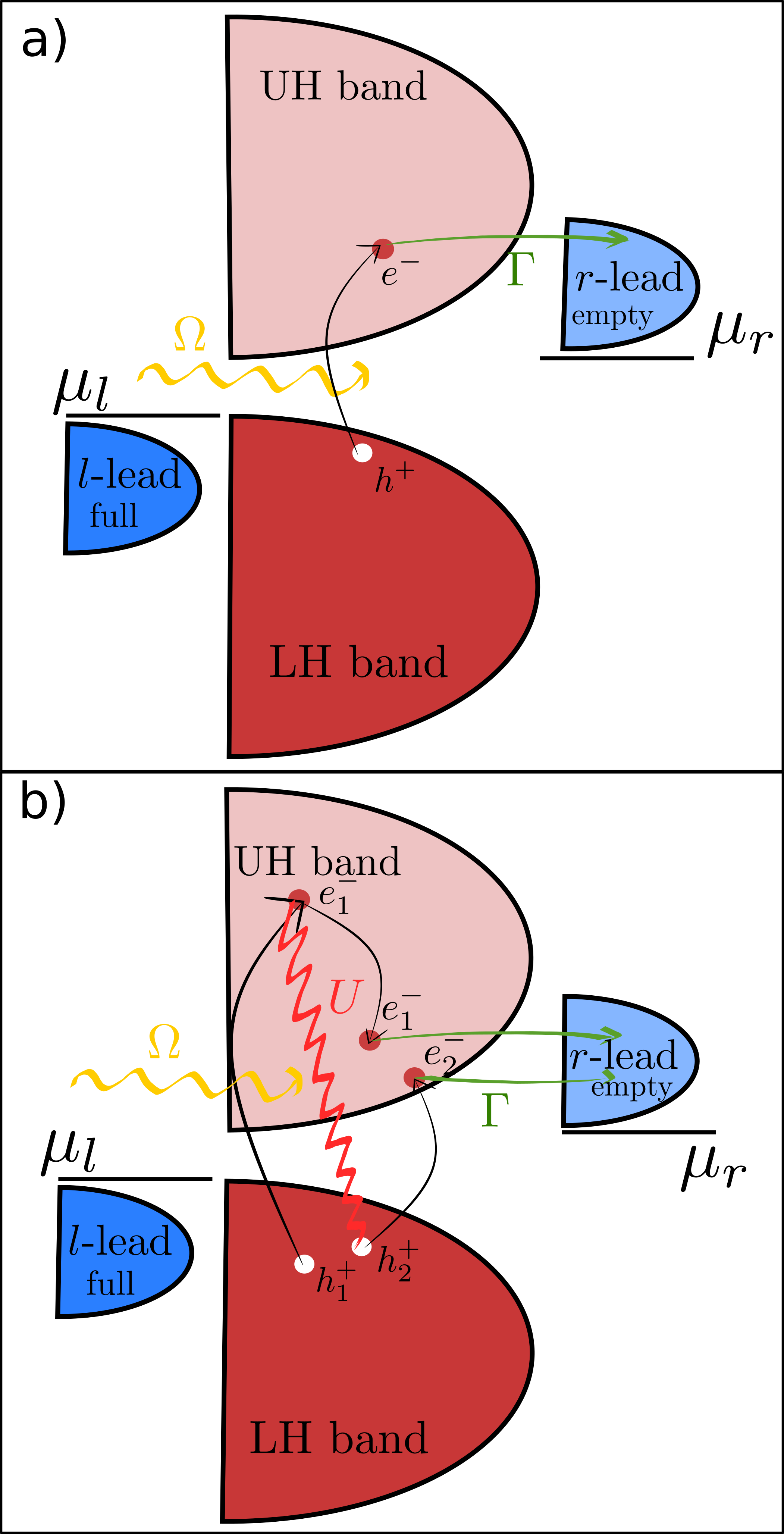}
\end{center}
 \caption{
A sketch of the 
energy distribution
 of the system studied  and an illustration of the dominant steady state photo-induced processes. The dark blue (light blue) region describe the full (empty) lead, while the lower and upper Hubbard bands of the central layer are marked in dark and light red, respectively.
Electromagnetic radiation with energy $\Omega$ (yellow) initially produces a particle-hole excitation (red-wiggly line).
For $\Delta_{ss}<\Omega<\Delta_{ss} + 2D$, panel (a)  an electron coming from the left (full) lead is photoexcited into the upper Hubbard band at energies such that it can directly escape into the right (empty) lead without further scattering processes. For $\Omega > 2 \Delta_{ss}$, panel (b), we have impact ionization:  First,  an electron hole pair is created with a high energy electron via photo absorption ($e^-_1,h^+_1$ in the figure). Since, the energy of the photoexcited electron is incompatible with states in the right lead it can not escape the correlated region. Thus, it can only contribute to the current if it can get rid of its excess energy. The simplest, and therefore dominant process is impact ionization, where $e^-_1$ scatters with a second electron from the lower band thereby exciting it over the steady state gap and creating a second e-h pair $e^-_2,h^+_2$. In the final state, both electrons and holes can now escape into the leads and contribute to the current. Notice that in the present steady state situation only processes that eventually recover the initial configuration are allowed.
}
 \label{fig:band}
\end{figure}

While the aforementioned studies have investigated the short- and medium-time dynamics of these model 
systems after a short electromagnetic pulse, in the present work we aim at studying a steady-state situation in 
which the system is under constant illumination and energy is continuously harvested by transferring electrons from a metallic lead at a lower  chemical potential into one at a higher chemical potential, see Fig.~\ref{fig:band}. We consider a purely electronic and highly simplified model for a Mott photovoltaic device consisting of a left/right lead and a correlated layer acting as photo-active region in between. To study the effect of impact ionization on the photovoltaic efficiency we investigate the photon-frequency resolved steady state photo-current, whereby the illumination is accounted for by coupling the correlated layer to an electric field oscillating with a single frequency.
Experimentally this would correspond to studying the photovoltaic effect with a laser in the laboratory.

The  energy  structure of the system is sketched in Fig.~\ref{fig:band}. 
In this special setup, one can distinguish between driving frequencies that support steady state current without 
the need of scattering (direct excitations Fig.~\ref{fig:band}a) and frequencies that require the production of an extra electron-hole excitation by  impact ionization Fig.~\ref{fig:band}b).
While the narrow bandwidth of the leads is certainly rather unconventional it is ideally suited for the detection of impact ionization, not only for theoretical means but also experimentally. Our main results, Fig.~\ref{fig:U12Current} below, show 
a steep increase of the current in the impact ionization case (by roughly a factor two in comparison to the case of direct excitations), accompanied by an increase in the double occupancy.

To corroborate the fact that direct excitations and impact ionization are the dominant processes in this steady state situation, we investigate the system also in a parameter region for which impact ionization is prohibited. Here, we find indeed a significant current only in the frequency regime of direct excitations.
For strong electric fields, however, we find possible signatures of, what we refer to as, higher order impact ionization processes.

In order to explore properties of the system in a steady state with the period associated with the frequency of the electric field, we employ a Floquet plus DMFT~\cite{sc.mo.02u,jo.fr.08,ts.ok.08} approach, whereby the recently introduced auxiliary master equation approach (AMEA)~\cite{ar.kn.13,do.nu.14,do.ga.15,do.so.17} is used as impurity solver. For simplicity,  we restrict  the self energy to be diagonal in the Floquet index. We find that this 
Floquet diagonal self-energy approximation (FDSA) is valid within the parameter range of interest by  testing it against a fully time-dependent  impurity solver. For this test, i.e.,  for being able to include the full time dependence, we employ the iterated perturbation theory (IPT).

This work is organized as follows: We introduce the model in Sec.~\ref{sec:Model} and outline the technical details related to the Floquet plus DMFT formalism 
used in this paper in Sec~\ref{sec:Method}. Some additional information and more elaborate discussions on some key points can be found in the Appendix together with a test of the validity of the FDSA. The main results and their discussion  are presented in Sec~\ref{sec:results} and our conclusions together with  an outlook on possible future investigations are presented in Sec.~\ref{sec: Conclusion and Outlook}.
 
\section{Model}\label{sec:Model}
To make for a first numerical study of the possible increase in photovoltaic efficiency due to impact ionization in the setting of a periodic drive, we work with the most basic model that captures only the key aspects of the physical situation. Since impact ionization is a purely electronic process, we work solely with electrons and neglect any other degrees of freedom. We want to note that in particular the coupling to lattice vibrations, including the polaron structure of the electronic quasiparticles, should be included in a more elaborate treatment as the latter are known to play a significant role in photo excited Mott systems~\cite{fi.ca.12,ok.mi.10,mi.na.08}\footnote{While the inclusion of phonons is possible within the present approach, it would take severe method development and numerical effort to do treat them non perturbatively.}. Other extensions of the model are discussed in our conclusions, see Sec.~\ref{sec: Conclusion and Outlook}.

\begin{figure}
 \includegraphics[width=0.9\columnwidth]{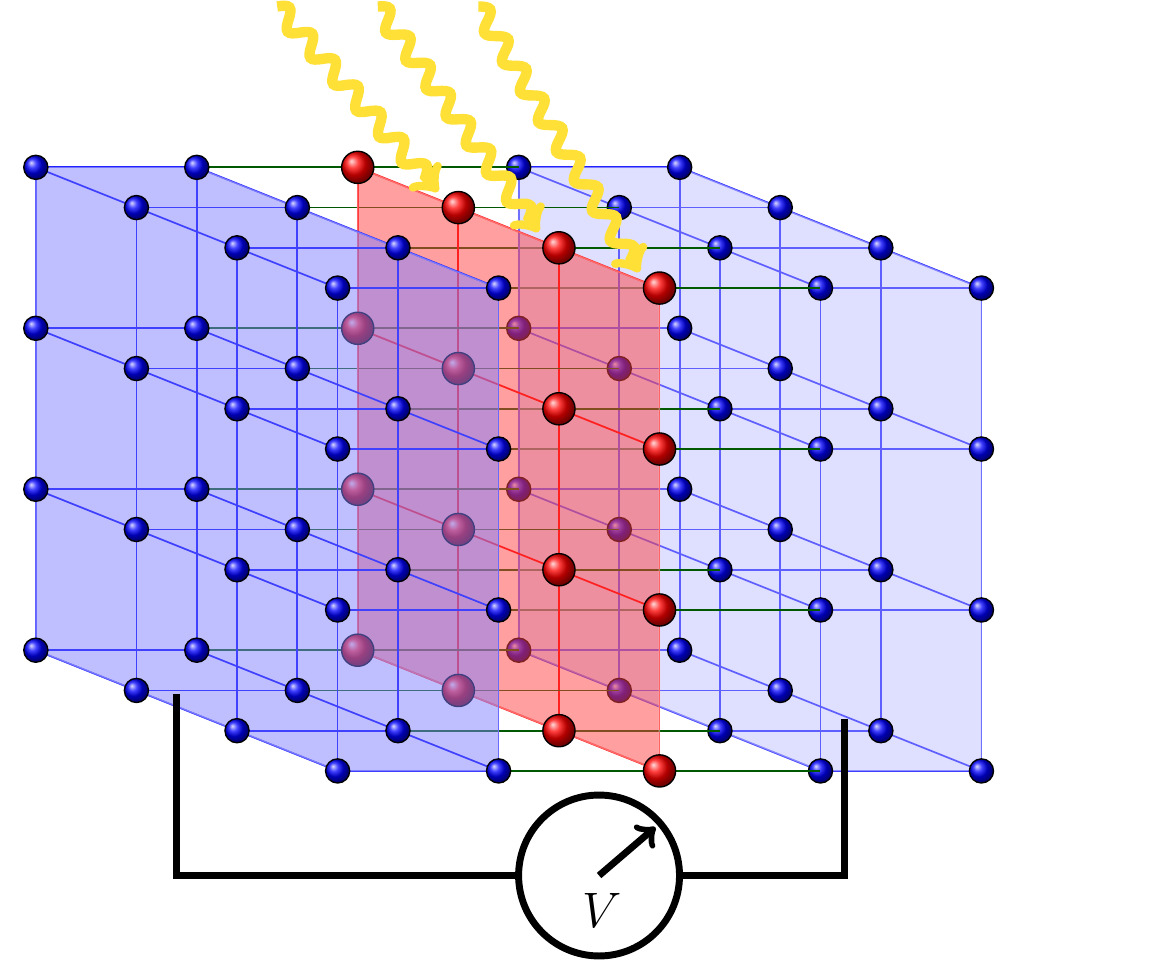}
 \caption{Sketch of the considered lattice system. The leads, Hubbard layer, monochromatic light
and the coupling between the lead and layer are illustrated in blue, red, yellow and green, respectively.}
 \label{fig:system sketch}
\end{figure}

In more detail, we consider a system consisting of a single band Hubbard layer connected on the two sides with metallic leads described by non-interacting tight-binding models. The central (Hubbard) layer is driven by a time-periodic, monochromatic and homogeneous electric field of frequency $\Omega$.
Fig.~\ref{fig:system sketch} shows a sketch of the lattice system. Its Hamiltonian  reads
\FloatBarrier
\begin{align}\label{eq:Hamiltonian}
&\hat{H}=\hat{H}_{\text{center}}(t) + \hat{H}_{\text{leads}} + \hat{H}_{\text{coupling}}\\
\notag&\hat{H}_{\text{center}}(t)=-\sum_{<ij>,\sigma}t_{ij}(t)c_{i,\sigma}^{\dagger}c_{j,\sigma} + \frac{U}{2}\sum_{i,\sigma}n_{i,\sigma}(n_{i,\bar{\sigma}} - 1)\\
\notag&\hat{H}_{\text{leads}}=\!\sum_{\gamma\in\{l,r\}}\!\Big(\!-t_{\gamma}\!\!\sum_{<ij>,\sigma}\!f_{\gamma_i,\sigma}^{\dagger}f_{\gamma_j,\sigma} + \sum_{i,\sigma}\epsilon_{\gamma}f_{\gamma_i,\sigma}^{\dagger}f_{\gamma_i,\sigma}\!\Big)\\
\notag &\hat{H}_{\text{coupling}}=\sum_{<i,j>\sigma} \left( V_lf_{l_i,\sigma}^{\dagger}c_{j,\sigma}+ V_rf_{r_i,\sigma}^{\dagger}c_{i,\sigma} + h.c. \right) \; .
\end{align}
Here, $c_{i,\sigma}^{\dagger}/c_{i,\sigma}$ denote creation/annihilation operators in the central layer and $n_{i,\sigma}=c_{i,\sigma}^{\dagger}c_{i,\sigma}$, while $f_{\gamma_i,\sigma}^{\dagger}/f_{\gamma_i,\sigma}$ refer to operators in the leads.
We consider a spatially uniform electric field along the diagonal direction of the central layer. By choosing the temporal gauge, where the scalar potential vanishes, the electric field is described by the vector potential $\boldsymbol{A}(t)=\boldsymbol{E}_0\cos(\Omega t)/\Omega$ resulting, according to the Peierls substitution rule, in a time-dependent hopping
\begin{equation}\label{eq:t_c(t)}
 t_{ij}(t)=t_ce^{-\frac{ie}{\hbar c}\boldsymbol{A}(t)(\boldsymbol{r}_j - \boldsymbol{r}_i)}.%\equiv t_ce^{i\alpha_{ij}\cos(\Omega t)}
\end{equation}
The units for the electric field are chosen such that the coefficient $\frac{e}{\hbar c}=1$. Moreover, we set the lattice spacing to unity and take $t_c$ as unit of energy throughout this work.

\subsection{Parameter setup}
\label{subsec:Parameters and Setup}
To study in particular the role of impact ionization on the steady state dynamics, we choose a very special and unconventional parameter setup that allows us to 
distinguish between regimes in which impact ionisation can take place or not as the external driving frequency is varied, hence enabling us to isolate the effect under study. To this end, we consider narrow leads with no overlap of their respective density of states and place the lead lower/upper in energy at the top/bottom of the lower/upper Hubbard band. In order to avoid the backflow of carriers into the source, the left (i.e., lower) lead is taken as completely filled, while the 
right one is empty.~\footnote{In a realistic system, this mechanism is provided by an intrinsic electric field, such as the one of p-n junctions or the polar field present in TMO heterostructures~\cite{as.bl.13}} Accordingly, the chemical potential $\mu_l/\mu_r$  lies just above/below the left/right band. Furthermore, we consider large  hybridization strengths $\Gamma = 2\pi|V|^2\rho(\omega=0)$,  with $V_L=V_R=V$ in Eq.~(\ref{eq:Hamiltonian}) and $\rho(\omega)$ denoting the density of states (DOS) of the leads, such that electrons with the correct energy escape quickly into the right lead\footnote{Strictly speaking this restricts only the right coupling but to stay at particle-hole symmetry we choose a symmetric hybridization.}, while it must be low enough not to dominate the dynamics in the central layer or to alter its corresponding DOS dramatically. Finally, we take moderate electric field strengths such that first order absorption processes dominate.

The situation is depicted in Fig.~\ref{fig:band}. In this setup, (particle-) current can only flow from the (filled) left lead into the (empty) right one and only when the system is externally driven. In the absence of scattering, the minimal frequency needed to drive a steady-state current is $\Omega\geq\Delta_{ss}$, where $\Delta_{ss}$ is the energy gap between the leads. Our aim here is not to provide a realistic model for photovoltaic applications, but rather to study and distinguish 
the different kind of process that can take place in a Mott photovoltaic in a steady state situation. As we will see below this setup is ideally suited to identify the existence of impact ionization in a steady state.
 
\section{Method}\label{sec:Method}

\subsection{Floquet Green's functions}
\label{subsec:Floquet GF}
To solve for the (periodic-)steady state properties of the system we work with the so called Floquet Green's function (GF)~\cite{fais.89,al.st.97,bran.97,br.ro.02,mart.03} formalism, which allows for the evaluation of the steady state current and spectral properties.
Here, every observable of the system, and thus also the single particle GF, is assumed to be periodic with the external driving frequency. Since a periodically driven system is inevitably out of equilibrium we work with non-equilibrium Keldysh GF's~\cite{kad.baym,schw.61,keld.65}. More precisely one defines the Floquet-Keldysh GF as
\begin{equation} \label{eq: Floquet GF}
 \underline{G}_{mn}(\omega)= \int dt_{\text{rel}}\frac{1}{\tau}\int_{-\frac{\tau}{2}}^{\frac{\tau}{2}} dt_{\text{avg}}e^{i(\omega+m\Omega)t - i(\omega+n\Omega)t'}\underline{G}(t,t')
\end{equation}
where $t_{\text{rel}}=t-t'$, $t_{\text{avg}}=(t+t')/2$, $m$ and $n$ denote the Floquet indices, and the underline indicates the Keldysh matrix structure
\begin{equation}\label{eq:Keldysh structure}
 \underline{G}=\begin{bmatrix}
 G^{R}&G^{K}\\
0&G^{A}
\end{bmatrix}
\end{equation}
with retarded, Keldysh and advanced component.
In Appendix \ref{app: Properties of Floquet GF} we mention some properties of Floquet GF that are of importance for the current work. Here, we just want to note that the time average (over one driving period) of a matrix in Floquet space
$\boldsymbol{X}(\omega)$
 is given by
\begin{equation}\label{eq: time av of Floquet object X}
 \bar{\boldsymbol{X}}(\omega) = \int_{-\frac{\tau}{2}}^{\frac{\tau}{2}} \frac{dt_{\text{avg}}}{\tau} X(\omega,t_{\text{avg}}) = X_{00}(\omega) \;.
 \end{equation}
 Here and in the following, we use boldface to indicate matrices in Floquet space (the only other boldface object is the wave vector $\boldsymbol{k}_{||}$, but there is no ambiguity).

\subsection{Dyson equation}
\label{subsec: Dyson equation}
The Dyson equation for the central Floquet lattice GF reads
\begin{equation} \label{eq:lattice GF}
 \underline{G}_{mn}^{-1}(\omega,\boldsymbol{k}_{||})= \underline{G}_{0,mn}^{-1}(\omega,\boldsymbol{k}_{||}) - \underline{\Sigma}_{mn}(\omega,\boldsymbol{k}_{||})
\end{equation}
where $\boldsymbol{k}_{||} = (k_x,k_y)$ 
is the crystal momentum  in the two translational invariant directions and  $\underline{X}_{mn}^{-1}$ denotes the $mn$ element of the inverse Floquet-Keldysh matrix.  The GF corresponding to the non-interacting part of the Hamiltonian Eq.(\ref{eq:Hamiltonian}) is given by
\begin{align} \label{eq:nonint lattice GF}
 \notag \underline{G}_{0,mn}^{-1}(\omega,\boldsymbol{k}_{||}) =& \underline{g}_{0,mn}^{-1}(\omega,\boldsymbol{k}_{||})\\
 &- \sum_{\gamma\in\{l,r\}}V^2_{\gamma}\underline{g}_{\gamma}(\omega_n,\boldsymbol{k}_{||})\delta_{mn}
\end{align}
with the shorthand $\omega_n\equiv \omega + n\Omega$ and
\begin{align} \label{eq:isolated lattice GF}
 \notag \left[\underline{\boldsymbol{g}}^{-1}_{0}(\omega,\boldsymbol{k}_{||})\right]_{mn}^{R}&=(\omega_n+i0^+ - \varepsilon_c)\delta_{mn} - \varepsilon_{mn}(\boldsymbol{k}_{||})\\
 \left[(\underline{\boldsymbol{g}}^{-1}_{0}\omega,\boldsymbol{k}_{||})\right]_{mn}^{K}&=0
\; .
\end{align}
Here, as usual in steady state, we can neglect the Keldysh component of the inverse Green's function of the layer, and
$\varepsilon_{mn}(\boldsymbol{k}_{||}) = \varepsilon_{mn}(k_x) + \varepsilon_{mn}(k_y)$. The Floquet dispersion relation in the presence of the periodic field described by Eq.~\eqref{eq:t_c(t)} is readily found to be~\cite{ts.ok.08}
\begin{align} \label{eq:Floquet disperion relation}
 \notag \varepsilon_{mn}(k)=&-t_c(-i)^{m-n}\\
 \times& \left[e^{ik}J_{m-n}\left(-\frac{E_0}{\Omega}\right) + e^{-ik}J_{m-n}\left(\frac{E_0}{\Omega}\right)\right]
\end{align}
with $J_n$ denoting the $n$-th order Bessel function of the first kind. The surface GF of the semi infinite, decoupled,  leads are given by
\begin{align}\label{eq:lead surface GF}
 \notag g_{\gamma}^R(\omega,\boldsymbol{k}_{||})=&\frac{\omega - \varepsilon_{\gamma}(\boldsymbol{k}_{||})}{2t_{\gamma}^2}\\
 \notag &-i\frac{\sqrt{4t_{\gamma}^2-(\omega-\varepsilon_{\gamma}(\boldsymbol{k}_{||}))^2}}{2t_{\gamma}^2}\\
g_{\gamma}^K(\omega,\boldsymbol{k}_{||})=&2i[1-2f_{\gamma}(\omega)]\Im g_{\gamma}^A(\omega,\boldsymbol{k}_{||})
\end{align}
where $f_{l/r}$ is the Fermi function for the $l/r$ lead and  $ \varepsilon_{\gamma}(\boldsymbol{k}_{||})=\varepsilon_{\gamma} -2t_{\gamma}(\cos(k_x) + \cos(k_y))$ is the usual dispersion relation for the simple cubic lattice.

\subsection{Time averaged Observables}
\label{subsec: Time averaged Observables}
In this work, we are interested in time averaged steady state observables. In particular, we focus on Green's functions, spectral functions and the current density through the correlated region. By virtue of Eq.~(\ref{eq: time av of Floquet object X}), the time average of a Floquet GF is readily obtained by picking out the $(0,0)$ component of the corresponding Floquet matrix. It is then natural to define the time averaged density of states or spectral function as
\begin{equation}\label{eq: def spectral function}
 A(\omega)=-\frac{1}{\pi}\Im G_{00}^{R}(\omega)
\end{equation}
which obeys the zeroth order spectral sum rule (normalization to unity) and reduces to the usual definition in the limit of no periodic driving.\\
To generalize a formula for a quantity that contains the product of GF's to the corresponding time averaged one in a Floquet system, care has to be taken as objects which commuted in the original formulation might not commute in the Floquet formalism. Thus, one has to be certain about the ordering in a given expression before applying the straightforward substitutions. For the case of the time averaged current a correctly ordered expression can be found in \cite{ha.ja,ar.kn.13} which, according to appendix \ref{app: Properties of Floquet GF}, is readily generalized to 
\begin{align}\label{eq: Current furmula}
 j_{L\rightarrow R}&=v^2\int_{-\frac{\Omega}{2}}^{\frac{\Omega}{2}}\frac{d\omega}{2\pi} \int_{\text{B.Z.}}\frac{d\boldsymbol{k}_{||}}{(2\pi)^2} \text{Re Tr} \boldsymbol{\mathcal{J}}\\
  &= v^2\int_{-\infty}^{+\infty}\frac{d\omega}{2\pi} \int_{\text{B.Z.}} \frac{d\boldsymbol{k}_{||}}{(2\pi)^2} \text{Re} \mathcal{J}_{00}
  \end{align}
where $\boldsymbol{\mathcal{J}}$ is a Floquet matrix given by
\begin{equation}
  \boldsymbol{\mathcal{J}}=\left[\boldsymbol{G}^{\text{R}}(\boldsymbol{g}_{l}^{\text{K}}-\boldsymbol{g}_{r}^{\text{K}})+\boldsymbol{G}^{\text{K}}(\boldsymbol{g}_{l}^{\text{A}}-\boldsymbol{g}_{r}^{\text{A}})\right]
\end{equation}
with upper case $\boldsymbol{G}$ denoting the lattice Floquet GF of the interacting region and lower case $\boldsymbol{g}_{l/r}$ refer to the two surface GFs of the decoupled leads. The two alternatives in Eq.~(\ref{eq: Current furmula}) can be used to check for consistency. They agree for  a sufficiently large Floquet matrix cutoff, and the agreement is a sign of convergence with respect to the cutoff. 

\subsection{Floquet DMFT}
\label{Floquet DMFT}
Since the correlated lattice problem cannot be solved exactly we have to resort to an approximate scheme for calculating the self-energy $\underline{\Sigma}_{mn}(\omega,\boldsymbol{k}_{||})$. To this end, we use dynamical mean field theory~\cite{me.vo.89,ge.ko.92,ge.ko.96} (DMFT) in its generalization to the periodically driven systems~\cite{sc.mo.02u,jo.fr.08,ts.ok.08}.  Within DMFT one neglects the $\boldsymbol{k}_{||}$-dependence of the self-energy, $\underline{\Sigma}_{mn}(\omega,\boldsymbol{k}_{||})\approx\underline{\Sigma}_{mn}(\omega)$, which allows one to calculate the approximate self-energy by the solution of a self consistently determined impurity-problem. In the following, we will provide only a very short description of the DMFT scheme and concentrate on the aspects due to the periodic time-dependence, for more details we refer the reader to the recent review on non-equilibrium DMFT\cite{ao.ts.14}.\\
With an initial guess for the self-energy $\underline{\boldsymbol{\Sigma}}(\omega)$, the first step of DMFT is  to obtain the local GF from the self-energy via the  $\boldsymbol{k}_{||}$-integrated Dyson equation for the lattice problem
\begin{equation} \label{eq:localgf}
 \underline{\boldsymbol{{G}}}_{\rm loc}(\omega)= \int_{B.Z.}\frac{d\boldsymbol{k}_{||}}{(2\pi)^2} \left[\underline{\boldsymbol{{G}}}_0^{-1}(\omega,\boldsymbol{k}_{||}) - \underline{\boldsymbol{\Sigma}}(\omega)\right]^{-1}
\end{equation}
The essential step is now the mapping onto an impurity problem. This step is achieved by considering the Dyson equation of the impurity model 
\begin{equation} \label{eq:imp dyson eq}
 \underline{\boldsymbol{G}}_{\text{imp}}^{-1}(\omega)=\underline{\boldsymbol{g}}_{\text{imp}}^{-1}(\omega) - \underline{\boldsymbol{\Delta}}(\omega) - \underline{\boldsymbol{\Sigma}}(\omega) \;.
\end{equation}
Here, $\underline{\boldsymbol{g}}_{\text{imp}}^{-1}(\omega)$ is the non-interaction Floquet-Keldysh GF of the impurity which is defined as in Eq.~(\ref{eq:isolated lattice GF}) but without $\varepsilon_{mn}(\boldsymbol{k}_{||})$.  %\kh{pls check}
Demanding equality of the local GF and the impurity GF, $\underline{\boldsymbol{{G}}}_{\rm loc}(\omega)\stackrel{!}{=}\underline{\boldsymbol{G}}_{\text{imp}}(\omega)$, we get the effective bath hybridization function
\begin{align}\label{eq:hybf}
\underline{\Delta}_{mn}(\omega)&=\underline{g}_{\text{imp},mn}^{-1}(\omega) - (\underline{G}^{-1}_{loc\ mn}(\omega) + \underline{\Sigma}_{mn}(\omega))
%&=\underline{g}_{\text{imp},mn}^{-1} -\underline{\mathcal{G}}_{0,mn}(\omega)
\end{align}
where we have explicitly reintroduced the Floquet indices (instead of the boldface matrices before) to emphasize that the corresponding impurity problem is now subject to a time-periodic driving. At this stage one inputs the obtained hybridization $\underline{\boldsymbol{\Delta}}(\omega)$ from Eq.~(\ref{eq:hybf}) to an impurity solver, obtains a new self-energy and iterates the steps above until convergence.

\subsubsection{Floquet-diagonal self energy}
\label{subsubsec: Time local Floquet DMFT}
As we have seen above, the Floquet DMFT equations lead to a periodically time-dependent bath for the impurity problem which makes the latter hard to solve. In the literature, the Floquet-DMFT impurity problem is treated with low order perturbative expansions that work directly in the time domain~\cite{sc.mo.02u,lu.kr.09,ts.ok.08,ts.ok.09,fran.13.qu,fran.13.po,le.pa.14,jo.fr.15}, for example IPT. While this is numerically possible to carry out quite easily, 
the drawbacks of such solvers is of course their limitation to certain parameter regimes, in the interaction and/or hybridization strength, There is also a limited error control when such solvers are  applied  to new situations where no benchmarks are available. 
In the present work, we  use instead the AMEA, a non-perturbative impurity solver, which is very accurate in  addressing steady-state situations in a wide range of parameters.~\cite{do.nu.14,do.ga.15}
When addressing photovoltaic effects, we are in the regime of weak periodic driving where first order photon absorption processes are dominant.
In this case, off diagonal Floquet terms are suppressed by a 
factor $\alpha \equiv t_c E_0/\Omega^2$\footnote{The $E_0/\Omega$ part is self evident as it appears explicitly in the Hamiltonian and as argument of the Bessel-functions. The parameter is also proportional to $t_c$ since as the latter goes to zero the explicit time dependence vanishes from the problem as well. The last factor of $\Omega$ in the denominator comes from a perturbative argument. Since the diagonal entries in the Floquet matrix are shifted energetically by a factor of $\Omega$, corrections around the diagonal limit, that is off-diagonal terms, are suppressed by precisely this factor, leaving us overall with the dimensionless quantity $\alpha = t_c E_0/\Omega^2$}.
Motivated by this, we  restrict the self energies to be diagonal in the  Floquet indices. This is in analogy with the original DMFT approximation where one takes the self energy as being diagonal in lattice indices. It allows the simplification of the Floquet impurity problem to a non-equilibrium steady-state one, albeit with time translation invariance. Of course, the periodic time dependence of the problem remains via the Floquet-index dependence of the noninteracting Green's function. In more technical terms, instead of solving an impurity problem with the hybridization Eq.~(\ref{eq:hybf}), we take the time average of the local GF, i.e. $\underline{G}_{loc\ 00}(\omega)$, and calculate from it a time translation invariant hybridization function, given by
\begin{equation}\label{eq:HybFDSEA}
 \underline{\Delta}_{\phys}(\omega)=\underline{g}_{\text{imp}}^{-1}(\omega) - \left[ \left(\underline{{G}}_{{\rm loc},00}\right)^{-1}(\omega) + \underline{\Sigma}(\omega)\right]
\end{equation}
The resulting steady state impurity problem is then solved to obtain $\underline{\Sigma}(\omega)$, and the Floquet self-energy is reconstructed as
\begin{equation} \label{eq:construct Floquet seflenergy}
 \underline{\Sigma}_{mn}(\omega) = \underline{\Sigma}(\omega+n\Omega)\delta_{mn} \, .
\end{equation}
This self-energy, plugged into Eq.~(\ref{eq:lattice GF}), in turn yields the full Floquet lattice GF. This approximation for the self-energy (FDSA) is of course an ad hoc one, since the AMEA solution of a periodic time-dependent problem would have been numerically too time consuming. To check its range of validity, we carry out a numerical test 
in Appendix \ref{app:Test_FDSEA} where we use IPT and find that it is very accurate 
in the parameter region we are interested in.

\subsubsection{Auxiliary Master equation approach Impurity solver}
\label{subsubsec:Impurity solvers}
For the sake of completeness, we briefly present the DMFT impurity solver used to obtain our results, namely the so-called auxiliary master equation approach
 (AMEA). For details we refer to our recent work~\cite{ar.kn.13,do.nu.14,do.ga.15,do.so.17}; for the IPT impurity solver see Appendix \ref{app:Test_FDSEA}. The key idea behind the DMFT-AMEA impurity solver, in close analogy with the exact diagonalization (ED)~\cite{ca.kr.94} in equilibrium, is to replace the bath of the original impurity problem obtained via the DMFT cycle and defined by the hybridization function (cf. \eqref{eq:HybFDSEA}) $\underline{\Delta}_{\phys}(\omega)$, with an auxiliary one described by a corresponding hybridization function $\underline{\Delta}_{\text{aux}}(\omega)$. 

In contrast to ED, this auxiliary bath is an open quantum system consisting of a finite number of sites  embedded into a markovian environment. One should, however, point out that  the dynamics at the impurity site are non-markovian. This auxiliary system, being finite, can be solved exactly by conventional Krylov-space methods~\cite{do.nu.14} or matrix-product states~\cite{do.ga.15},  and the self-energy at the impurity site can be extracted. The only approximation entering the approach comes from the difference between the $\underline{\Delta}_{\phys}(\omega)$, and the auxiliary one $\underline{\Delta}_{\text{aux}}(\omega)$ provided by the non-markovian open system. The parameters of the latter 
 are determined by a fit requiring that  $\underline{\Delta}_{\phys}(\omega)\approx\underline{\Delta}_{\text{aux}}(\omega)$ as close as possible. 
This mapping  shows an exponential convergence~\cite{do.so.17} with respect to the number of bath sites in the auxiliary system, allowing quick convergence of results. This has to be, of course, confronted with the exponential growth in the numerical effort to solve the open quantum system.

\section{Results and Discussion}
\label{sec:results}
We investigate our model in two different regimes, $R_a$ and $R_b$ characterized by different values of the Hubbard gap. We show results for time-averaged quantities, namely
the current density, Eq.~(\ref{eq: Current furmula}) and (time-averaged-) spectral/Green's functions, Eq.~(\ref{eq: def spectral function}), as the driving frequency is varied.
For regime $R_a$, we check our method and implementation by considering the model in a parameter regime incompatible with impact ionization where the result for the current is severely limited by simple arguments as discussed below. The second case, $R_b$, allows us to directly study the effect of impact ionization. 
The corresponding parameters are summarized in table (\ref{tab:parameters}).

\begin{table}
 {
\newcommand{\mc}[3]{\multicolumn{#1}{#2}{#3}}
\begin{center}
\begin{tabular}{|c|c|c|c|ccc|c|c}\hline
× & $U$ & $E_0$ & $t_{l/r}$ & \mc{1}{l|}{$\varepsilon_{l/r}$} & \mc{1}{l|}{$D$} & $\Delta_{SS}$ & $V_{l/r}$ & \mc{1}{l|}{$\Gamma_{l/r}$}\\\hline
× & \mc{2}{c|}{$H_c$} & \mc{4}{c|}{$H_{leads}$} & \mc{2}{c|}{$H_{coupling}$}\\\hline
$R_a$ & $12$ & $2$ & $1/6$ & \mc{1}{c|}{$\pm 3$} & \mc{1}{c|}{$2$} & $4$ & $0.8$ & \mc{1}{c|}{$9.6$}\\\hline
$R_b$ & $30$ & $12$ & $1/6$ & \mc{1}{c|}{$\pm 12$} & \mc{1}{c|}{$2$} & $22$ & $0.8$ & \mc{1}{c|}{$9.6$}\\\hline
 \end{tabular}
 \end{center}
}
\caption{The parameters for the two considered regimes in accordance with Eq.~(\ref{eq:Hamiltonian}) where $t_c$ serves as unit of energy. Additionally, we work roughly at room temperature by setting $k_bT=0.02$ in all calculations. Please recall, $\Gamma=2\pi|V|^2\rho(\omega=0)$, $D$ denotes the bandwidth of the leads.}
\label{tab:parameters}
\end{table}

\subsection{General considerations: Direct excitations versus impact ionization}
\label{dirim}
In order to yield impact ionization processes the  bandwidth of the upper Hubbard band of the central correlated layer must be larger than twice the  gap. 
Only if this is the case, the photoexcited electron (or hole) in the upper (lower) Hubbard band has enough excess energy to excite an additional electron across the band gap, i.e., to create a second electron-hole pair.Before presenting our actual results, we would like to first discuss the physical processes that we expect upon increasing the photon frequency $\Omega$. For the following discussion, we will assume that only first order light absorption processes are possible.

First,  for $\Omega<\Delta_{ss}$, we have a situation  with no  current, provided the central DOS has a true gap as in Fig.~\ref{fig:band}.
In the case of a partial gap, as we have in Fig.~\ref{fig:U12Current}c, the current should be suppressed, as multiple absorption processes will be needed to overcome the steady state gap.

For larger frequencies $\Delta_{ss}<\Omega<\Delta_{ss} + 2D$ electrons coming from the left lead are photo-excited to an energy within the bandwidth of the right lead and can directly escape, without the need of further scattering. We  refer to such processes, that are illustrated in Fig.~\ref{fig:band} a), as direct excitations. In addition, the strong hybridization with the leads ensures that a charge carrier with the right energy quickly leaves the central layer to the other side. Note that in a non-interacting model, say a band insulator for the central layer, this would be the only $\Omega$ regime with non-zero current.

 Next, we could have, in principle, an intermediate frequency region where the driving frequency is too big for a direct excitation but too low for steady state impact ionization, $\Delta_{ss} + 2D<\Omega<2\Delta_{ss}$. However, we will work with parameters such that  $\Delta_{ss} = 2D$ and therefore this region is absent. 

Finally, for $\Omega>2\Delta_{ss}$ we enter the regime of impact ionization, illustrated in Fig.~\ref{fig:band}b). In this regime each absorbed photon produces two carriers, so that one should expect a current up to twice as large for fixed absorption rate. The abrupt increase of the current at a given frequency hence signals the presence of impact ionization. In our model which neglects phonons and considers the  steady state, intraband scattering is not effective in allowing high-energy electrons to dissipate energy in the upper Hubbard band as we argue in  Appendix \ref{app:intrabandscattering}. For this reason,  we can be certain that the current observed for  $\Omega>2\Delta_{ss}$ can only be due to interband scattering from which steady state impact ionization should be the dominant one.

\subsection{Current and spectral functions}
\label{subsec: Current and spectral functions}

Another aspect that we need to consider  
before presenting the results for the current density,  is 
that a certain background current is intrinsic within the AMEA approach in the presence of spectral gaps or a band edge.
This is due to the fact that the fit to the hybridization function cannot go to zero abruptly when a gap is present, since   sharp features are hard to resolve when fitting with smooth functions. We have therefore estimated a ``background current'' to be removed from the results presented in Figs.(\ref{fig:U12Current}) (\ref{fig:U30Current}) and (\ref{fig:U30DOS}b). For example, for large $\Omega$ the current should go to zero in all cases, as there are no final states available for photoexcitation, but instead it settles at a finite value. 

This value is the same as the one we get when the external drive is switched off. The background that is estimated by switching off the electric field is
 indicated by the wiggled lines in Figs.(\ref{fig:U12Current}) (\ref{fig:U30Current}) as well as (\ref{fig:U30DOS}b) and agrees with the residual AMEA current in the region where no current is expected because of large (small) $\Omega$. Unfortunately, the background current is not always independent of frequency, so that it is sometimes hard to identify it. This occurs, for example, when the accuracy of the AMEA fit changes considerably in different frequency regions due to a crossover to different DMFT solutions, as it is the case in Fig.~\ref{fig:U30DOS}.

\subsubsection{Regime without impact ionization}
\label{subsubsec:Regime without impact ionization}
\begin{figure}[h]
 \includegraphics[width=0.9\columnwidth]{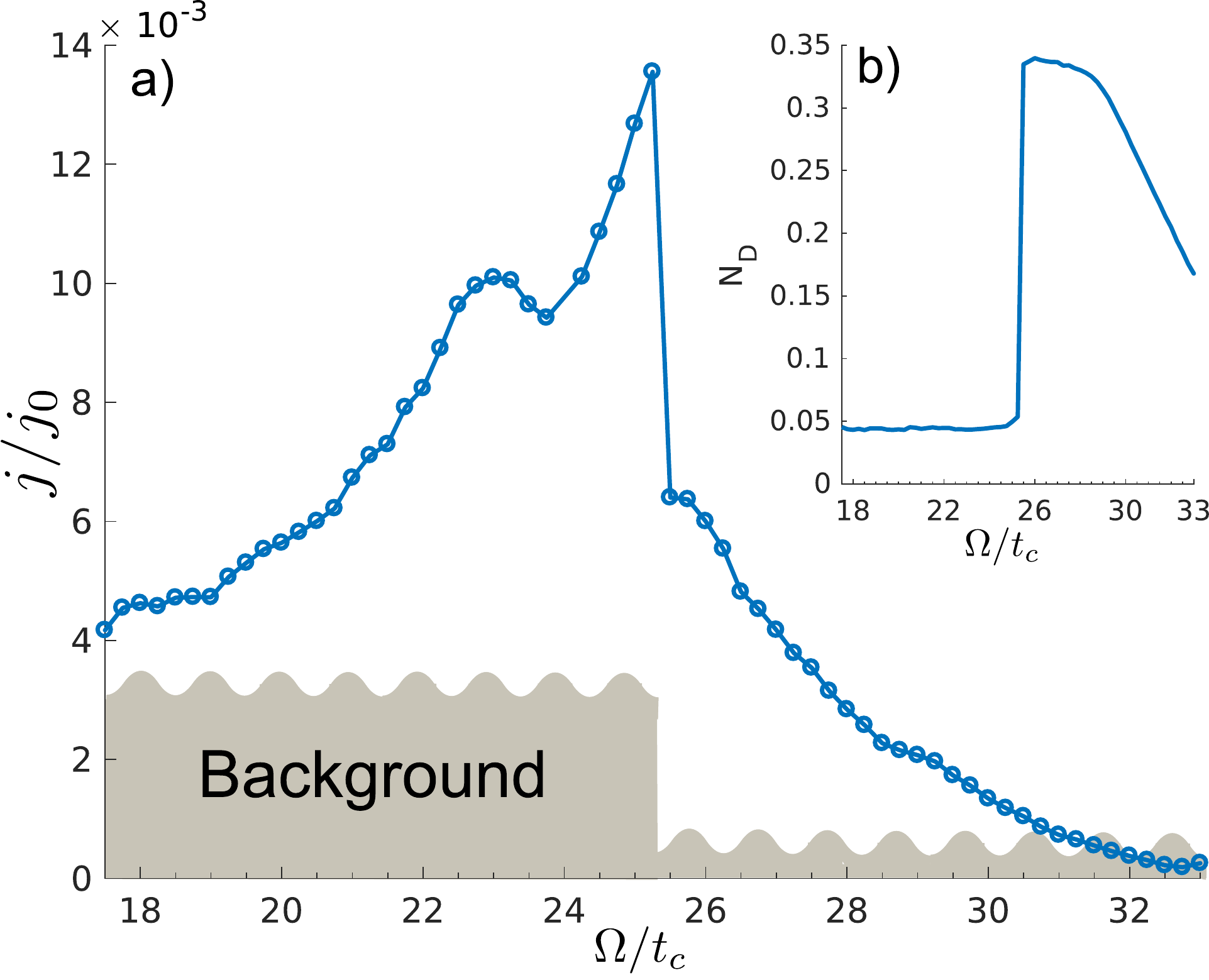}
 \caption{a) Time averaged current density and b) double occupancy as function of driving frequency for a situation with a gap larger than the bandwidth so that impact ionization is impossible. Parameters as in $R_a$ in table~\ref{tab:parameters}.}
 \label{fig:U30Current}
\end{figure}

We start by considering the system in the large gap regime, $R_a$ in table \ref{tab:parameters}. In this case,  for frequencies beyond $\Delta_{ss}+2D$ (see Sec.~\ref{dirim}), no current should be expected, as high energy doublons are trapped in the higher Hubbard band and cannot dissipate their energy (see also Appendix \ref{app:intrabandscattering}). The corresponding spectral function and filling is shown in Fig.~\ref{fig:U30DOS}. Since impact ionization is not allowed there should be a substantial current density only in the regime of direct excitations $22<\Omega<26$. The corresponding plot in Fig.~\ref{fig:U30Current} indeed  shows 
a current above the background in this expected frequency region and a clear double peak structure. The latter feature is a simple DOS effect consistent with the picture of direct excitations\footnote{A double peak structure already appears when considering only the density of states at the initial and final state of the photo-excitation in the lower/upper Hubbard band, i.e. $A(\omega)A(\omega+\Omega$, where the initial states are restricted to the support of the left lead.}. The fact that the current shows substantial broadening on the edges of the expected frequency region is due to the above mentioned limited resolution of the AMEA fitting procedure. Notice that within the FDSA the impurity solver has knowledge about the frequency through the DMFT self-consistency, which, for example, affects the occupation of the upper Hubbard band. Further, we can see that in the region $\Omega>25$ the background current is much smaller than at lower frequency. This is because the spectral situation changes around that point, since for $\Omega>24$ electrons can get excited from the lower lead directly in the trapping region above the right lead, leading to an accumulation of high energy doublons. In Fig.~\ref{fig:U30DOS} this is reflected in a considerable  filling of high frequency states. The different situation for $\Omega>25$ in turn allows for a better DMFT fit. This is the reason why also the background level in Fig.~\ref{fig:U30Current}  changes around  $\Omega=25$.\footnote{Even if the value of the cost function in the AMEA fit for the two situations would be the same, the background current can still be very different as the resolution of the (lead-) band edges might differ substantially.}. In summary, the current in Fig.~\ref{fig:U30Current} is consistent, within the limited accuracy of the present method, with the physical expectations and we can be confident that our approach captures the relevant physics.

\subsubsection{Regime with impact ionization}
\label{subsubsec:Regime with impact ionization}
\begin{figure}[h]
 \includegraphics[width=0.9\columnwidth]{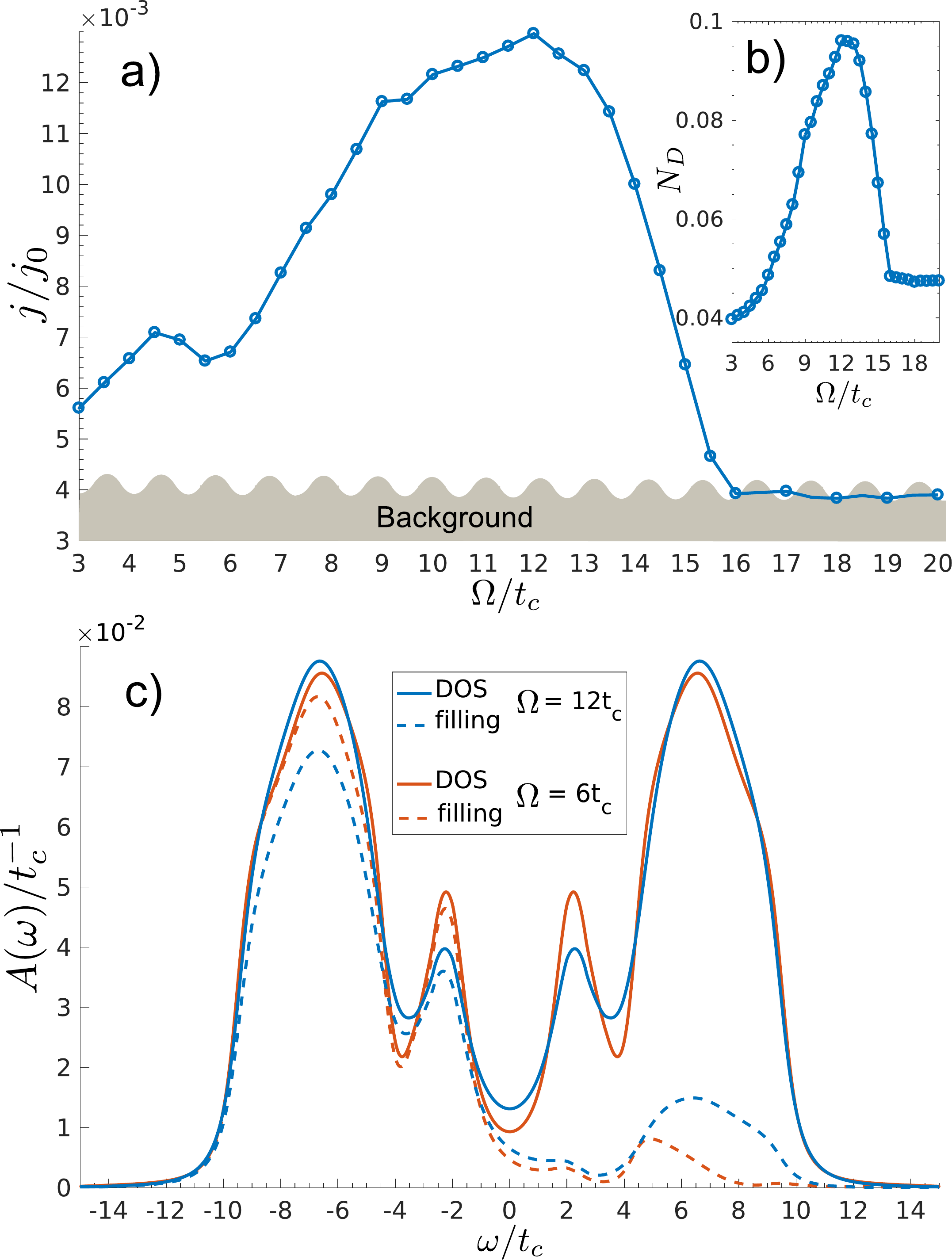}
 \caption{a) Time averaged current density
(in units of $j_0=e\hbar/t_ca^2$)
 and  b) double occupancy as a function of driving frequency for parameters compatible with impact ionization, $R_b$ in table~\ref{tab:parameters}. 
The current and the double occupancy show the same behavior in the region of the main peak as expected for impact ionization processes. c) Time averaged spectral function $A(\omega)$  and  corresponding filling $A(\omega) f(\omega)$ \protect \footnote{Here, $f(\omega)$ denotes the non-equilibrium distribution function.} for representative driving frequencies for direct excitations ($\Omega=6$) and impact ionization ($\Omega=12$).
}
 \label{fig:U12Current}
\end{figure}

Having established the correctness of our approach in the simple but nontrivial large gap case, we now consider the second set of parameters $R_b$ in table \ref{tab:parameters}. In this case, there is the possibility of impact ionization as sketched in Fig.~\ref{fig:band} and discussed in Sec.~\ref{dirim}.
The current density as a function of  $\Omega$ is plotted in Fig.~\ref{fig:U12Current} where we can identify three domains separated by smooth transitions: The region around the first maximum at $\Omega  \approx 4.5 $ corresponds to direct excitations, which are allowed for $4<\Omega<8$, see Sec.~\ref{dirim}. The second domain, hosting the main peak, is consistent with impact ionization, $8<\Omega<16$\footnote{It should be noted that in the steady state situation the $\Omega$ region where (steady state) impact ionization is possible, is more restrictive than in the case of an isolated Hubbard system after a photo-excitation, see Sec.~\ref{par:Higher order steady state impact ionization}}. The  maximum current for these frequencies is roughly twice as large as the one in the region of direct excitations. This corroborates the fact that a single photo absorption produces two charge carriers in the frequency region for impact ionization. 
The current for large driving frequencies $\Omega>16$ is just the  background current mentioned above.

From Fig.~\ref{fig:U12Current}b and c, we can see that going from a situation of direct excitations to one with impact ionization the double occupation and, in particular,
the occupation of high energy states (high energy doublons) compatible with impact ionization increases. 
This is consistent with our interpretation, that the second peak in Fig.~\ref{fig:U12Current}a originates from the latter process. Also in experiment such an abrupt doubling of the current could serve as clear indication of impact ionization. For example,
this can be used to detect impact ionization in semiconductor quantum dots~\cite{pi.ul.09}, cf. Ref. \onlinecite{se.lu.11}.

\subsubsection{Instability to multiparticle impact ionization processes}
\label{subsubsec:DMFTinstabilities}
Coming back to the large gap case, $R_a$, an intriguing finding of our analysis is that we actually find two distinct  non-equilibrium solutions for large driving frequencies. That is, for  $\Omega \gtrsim 35$ there are two DMFT  solutions (depending on the initial DMFT self-energy), see inset of  Fig.~\ref{fig:U30DOS}.
While this makes the behavior of the current in this region inconclusive, it is instructive to study the spectral properties of these two solutions, which we plot 
in Fig.~\ref{fig:U30DOS}. In the first solution there is an accumulation of high-energy doublons in the upper Hubbard band and a suppressed  current. The second solution is not showing this charge accumulation and supports a substantial steady state current above the background for $33\lesssim\Omega\lesssim 38$ hinting at a possible dissipation channel. This second solution gets unstable at lower driving frequencies below $\Omega \approx 35$. On the other hand, this second solution
becomes more stable at larger values of the electric field (not shown). We speculate that this behavior may be due to the occurrence of higher order impact ionization as explained in the following:

\begin{figure}[h]
 \includegraphics[width=0.9\columnwidth]{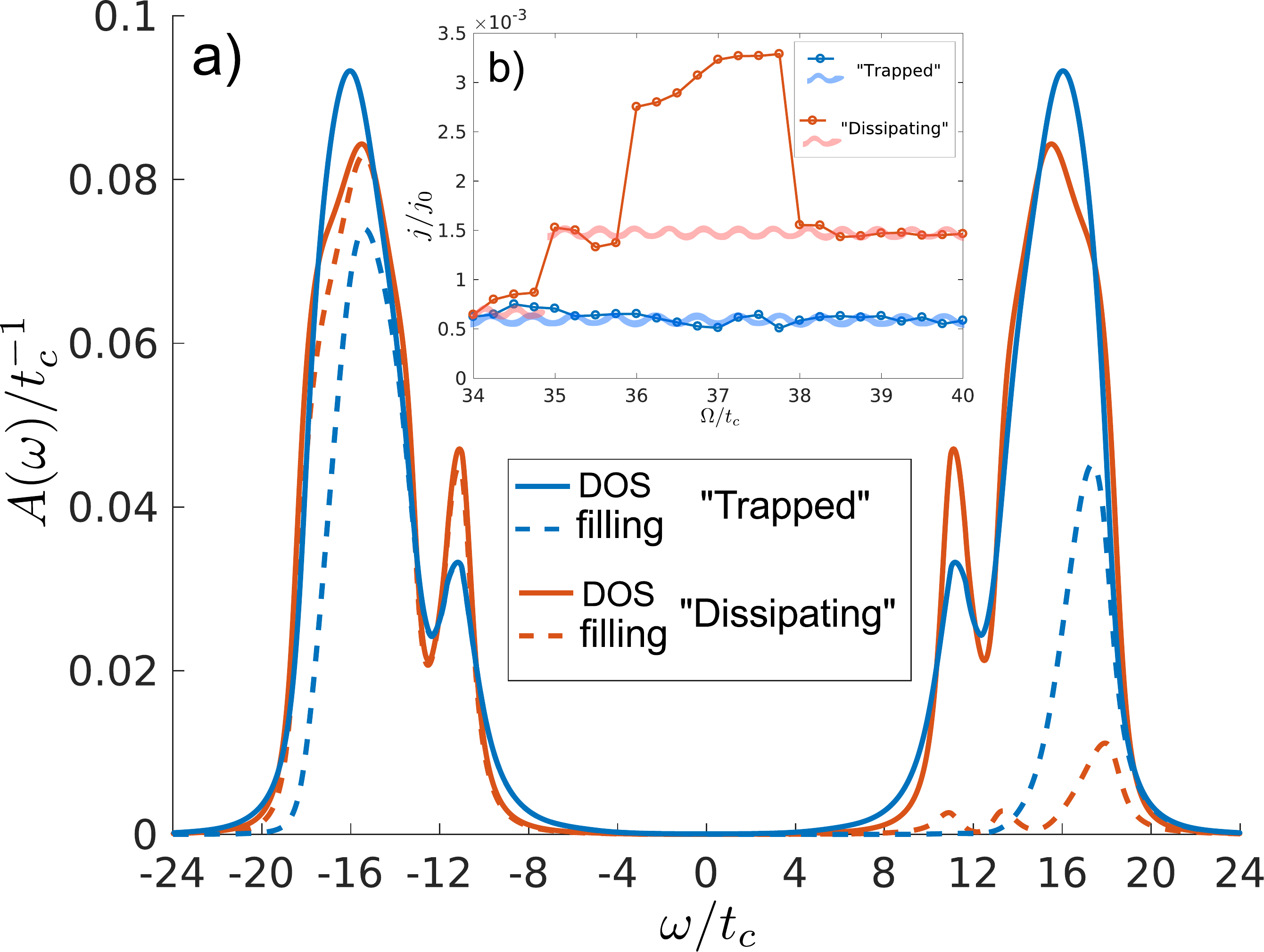}
 \caption{a) Time averaged spectral function and filling at $\Omega=37$ b) time averaged current density as function of driving frequency for the two solutions discussed in Sec.~\ref{subsubsec:DMFTinstabilities}. Parameter set $R_a$ as in Fig.~\ref{fig:U30Current}}.
 \label{fig:U30DOS}
\end{figure}

\paragraph{Higher order steady state impact ionization: }
\label{par:Higher order steady state impact ionization}
As already mentioned, a steady state process must be such that all the energy going into the system is dissipated again. For the present situation we have one \textit{source} of energy, the driving field characterized by $\Omega$, and one \textit{drain} of energy, namely the possible energy differences of the leads given by $\phi\equiv\varepsilon_{r} - \varepsilon_{l} \pm D$, i.e., for the parameters considered $22<\phi<26$. Therefore, we must have
\begin{equation}\label{eq: ss energy consvervation}
m\Omega\stackrel{!}{=}n\phi, \text{with } n\geq m
\end{equation}
in order for a process to be allowed. In view of Eq.~(\ref{eq: ss energy consvervation}), direct excitations correspond to $m=n=1$ while ordinary impact ionization is obtained when $n=2, m=1$. The next process is $n=3,m=2$. Here  two ``photons'' excite three electrons over the gap. Of course, this process is higher order, so that it will be visible only at large driving intensities. At this point,  one can argue that the second metastable DMFT solution mentioned above is related to the latter process as it falls  in the 
corresponding regime with $m=3,n=2$, i.e.,  $33<\Omega_{2m=3n}<39$.  This interpretation is supported by the fact that the second solution gets more stable at higher light intensities. However, we stress that the interpretation is still very speculative as, due to several reasons connected with our approach, this frequency region is difficult to address.

\section{Conclusion and Outlook}\label{sec: Conclusion and Outlook}
We have investigated the effect of impact ionization on the  current density through a periodically driven Mott insulator in the (periodic-) steady state using a simplified model for a Mott photovoltaic system. As a function of driving frequency we identify a crossover between a regime of direct excitations into one in which impact ionization take place and the current is substantially enhanced.

We also consider the deep Mott regime, where the Hubbard gap is larger than the bandwidth such that impact ionization is not possible. Here, we find hints for competing non-equilibrium phases of the system for larger driving frequencies. We give a possible interpretation of this behavior in terms of higher order impact ionization processes where multiple photo-excitations together with higher order interband scattering open a  dissipation channel supporting a non-vanishing current. The present work addresses a simplified model to study photovoltaic processes in a Mott solar cell but can be generalized in several directions to make for a more realistic modeling of actual solar cells. For instance, realistic metallic leads have typically a wide band and are only partially filled. Instead we use narrow bands in the leads which are optimally suited to separate impact ionization from other processes. Such narrow lead bands can be realized in organic crystals which have a small hopping amplitude or in materials with strong spin-orbit coupling which splits the bandstructure into several subbands. The extreme situation of zero bandwidth can be realized by bridging the photoactive region
through molecules  to the leads. Indeed this approach is employed for semiconductor quantum dots \cite{Wa.mc.13}, e.g. with the purpose to extract hot, photoexcited carriers from the quantum dot \cite{ti.ke.10,ca.wa.16}. Moreover, for solar cells based on oxide heterostructures the correlated region should consist of multiple layers making for the possibility to model an electric field gradient which separates  electrons and holes. On top of this, one should account for electron-phonon interactions and also long-range Coulomb forces to address bound excitons.
 
As discussed in Sec.~\ref{subsubsec: Time local Floquet DMFT}, in this work we have restricted to a time-translation-invariant hybridization function. In principle, the solution of the full time-periodic (Floquet) impurity problem can also be obtained within AMEA, and it would therefore be interesting to address the effects of a time-dependent self-energy beyond the FDSA. This would, however, be numerically rather expensive and relevant only in the case of strong driving which is not relevant for solar cell applications.

\begin{acknowledgments}

We would like to thank 
D. Fugger,
I. Titvinidze, 
W. von der Linden, 
M. Eckstein,
H.G. Evertz and F. Maislinger
for fruitful discussions.
This work was
partially supported by the Austrian Science Fund (FWF)  within 
Projects  P26508 and F41 (SFB ViCoM),  as well as NaWi Graz.
The calculations were partly performed on the D-Cluster Graz 
and on the VSC-3 cluster Vienna

\end{acknowledgments}

\appendix

\section{Properties of Floquet GF}
\label{app: Properties of Floquet GF}
First, the Floquet transform, defined through Eq.~(\ref{eq: Floquet GF}), for a GF that depends only on the time difference $G(t,t')=G(t-t')=G(t_{\text{rel}})$ leads to a diagonal Floquet matrix
\begin{equation}\label{eq:Floquet GF diag property}
\int dt_{\text{rel}}\int_{-\frac{\tau}{2}}^{\frac{\tau}{2}}\frac{dt_{\text{avg}}}{\tau}e^{i\omega_mt - i\omega_nt'}G(t_{\text{rel}})=G(\omega_n)\delta_{mn}
\end{equation}
with $\omega_n\equiv \omega+n\Omega$. The Floquet matrix entries are then not independent, but we have
\begin{equation} \label{eq: Floquet matrix relations}
 G_{mn}(\omega)=G_{m-n,0}(\omega+n\Omega) \;.
\end{equation}
For the important case of the equal time correlation function, we get
\begin{equation}
 G(t=t')=\sum_{m,n}e^{-i(m-n)\Omega t}\int_{-\frac{\Omega}{2}}^{\frac{\Omega}{2}}\frac{d\omega}{2\pi}G_{mn}(\omega) \; ,
\end{equation}
and for its time average
\begin{align}
 \notag \int_{-\frac{\tau}{2}}^{\frac{\tau}{2}}\frac{d\tau}{\tau}G(t=t')&=\sum_{n}\int_{-\frac{\Omega}{2}}^{\frac{\Omega}{2}}\frac{d\omega}{2\pi}G_{nn}(\omega)\\
 &=\int_{-\infty}^{+\infty}\frac{d\omega}{2\pi}G_{00}(\omega) \; .
\end{align}
Second, 
we want to note that the time average of a quantity is always encoded in the diagonal contributions of its corresponding Floquet matrix and thus by virtue of Eq.~(\ref{eq: Floquet matrix relations}) contained in the $m=n=0$ component alone.
Finally we want to mention the most appealing aspect of Floquet GF's, namely that a convolution in time is mapped to the multiplication of the corresponding Floquet GF's which leads to an algebraic Dyson equation in frequency.

\section{Intraband scattering processes in steady state}
\label{app:intrabandscattering}
It is important to stress that in our steady-state setup and in the absence of other inelastic scattering processes besides electron-electron interaction, high-energy doublons lying above the upper edge of the upper lead cannot easily dissipate their energy 
via intraband processes  so as to  be able to exit via the drain lead. Indeed, if a particle (A) loses a certain energy $\Delta$, a second particle (B) must, at the same time, gain that same amount. For the case in which (B) is in the upper Hubbard band, this would produce an accumulation of particles in the high-energy region of the band, as observed in Fig.~\ref{fig:U30DOS}. However, in  a stationary situation the rate of particles flowing  into this energy region must be equal to their outflow. For this reason,  these high energy particles must also find a channel to dissipate their energy, but again only electron scattering is available. For the case in which (B) is in the
lower Hubbard band, this process would produce an accumulation of particles in the upper part and a depletion in the lower part of the band, and we have the same situation as above.

The only possibility is that the energy $\Delta$ is large enough so that particle B is excited across the lead gap, i.e., impact ionization. In the situation of section \ref{subsubsec:Regime without impact ionization}, however, the central band width is smaller than the gap, so that this is possible only within multi-scattering and absorption processes. This statement may sound counterintuitive, and in fact in a realistic situation, acoustic phonons will carry away the energy excess. Therefore, the present results are valid for the case that these scattering processes are faster than the electron-phonon dissipation rate. In principle, also magnons are  relevant for energy dissipation~\cite{le.pr.13,ec.we.14}. However, since magnons consist themselves of  electronic excitations, electron-magnon scattering is  simply another form of electron scattering and the argument above remains valid. In a steady state situation we are not able to steadily transfer energy and excite  magnons in the central Hubbard layer.

\section{Test of the validity of the Floquet-diagonal self-energy}
\label{app:Test_FDSEA}
The goal of this appendix is to test the range of validity for the Floquet-diagonal self-energy approximation (FDSA).
The AMEA impurity solver, while being numerically controlled, is computationally expensive when carrying out a real time evolution. Hence, for this analysis we employ IPT as an impurity solver, which is much cheaper. Specifically, we compare time averaged observables from calculations with and without the FDSA.

\begin{figure}[h]
 \includegraphics[width=0.9\columnwidth]{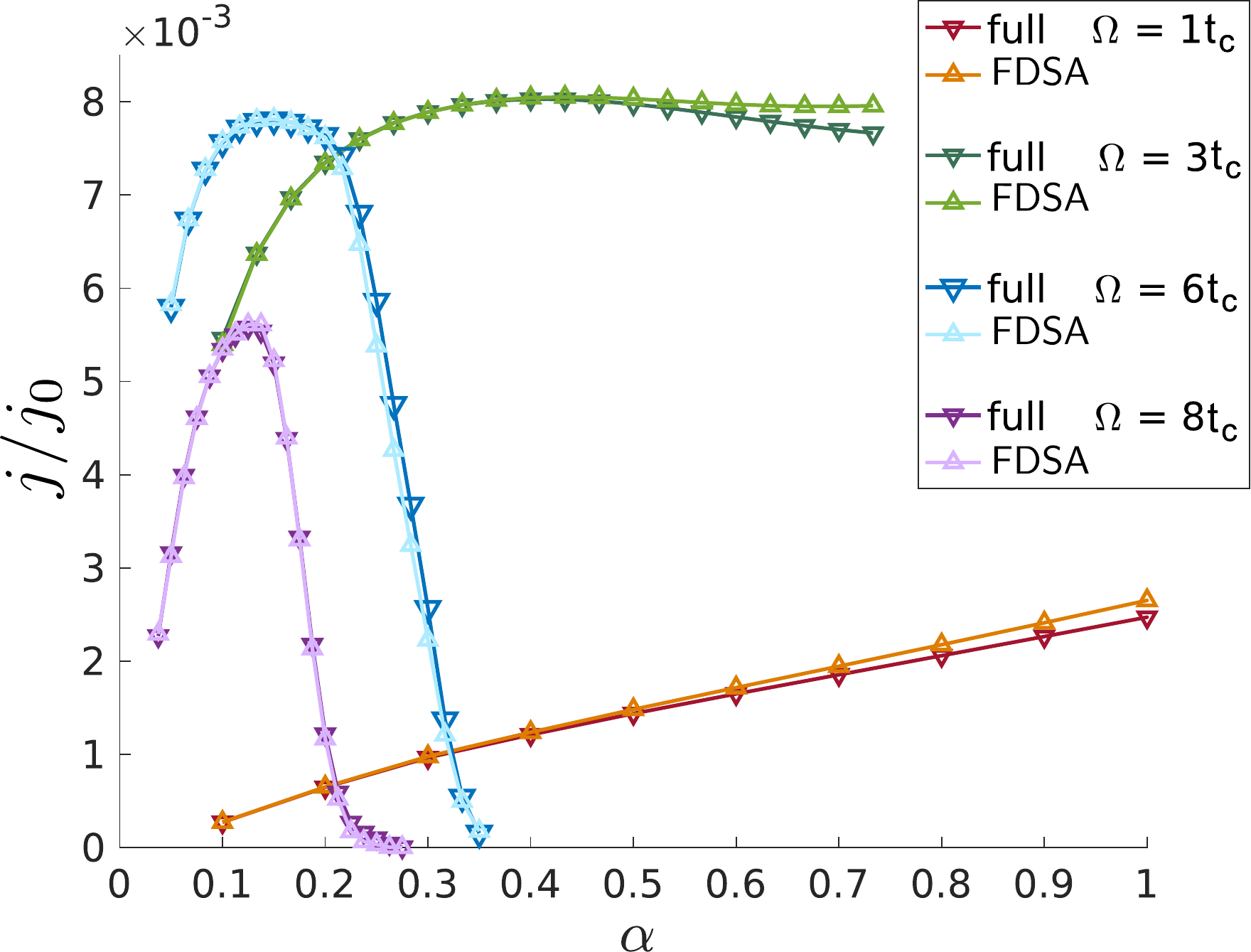}
 \caption{Test of the FDSA,  see Sec.(\ref{subsubsec: Time local Floquet DMFT}). Time averaged current as function of the effective coupling constant $\alpha\equiv t_c E_0/\Omega^2$ for different driving frequencies obtained with IPT. The lines labeled with 
``FDSA'' correspond to data computed with the FDSA, while the label ``full'' refers to the solution with the full Floquet impurity problem. For $\alpha\lesssim 1/2$, the FDSA approximation is very reliable.}
 \label{fig:IPT Current}
\end{figure}

\begin{figure}[h]
 \includegraphics[width=0.9\columnwidth]{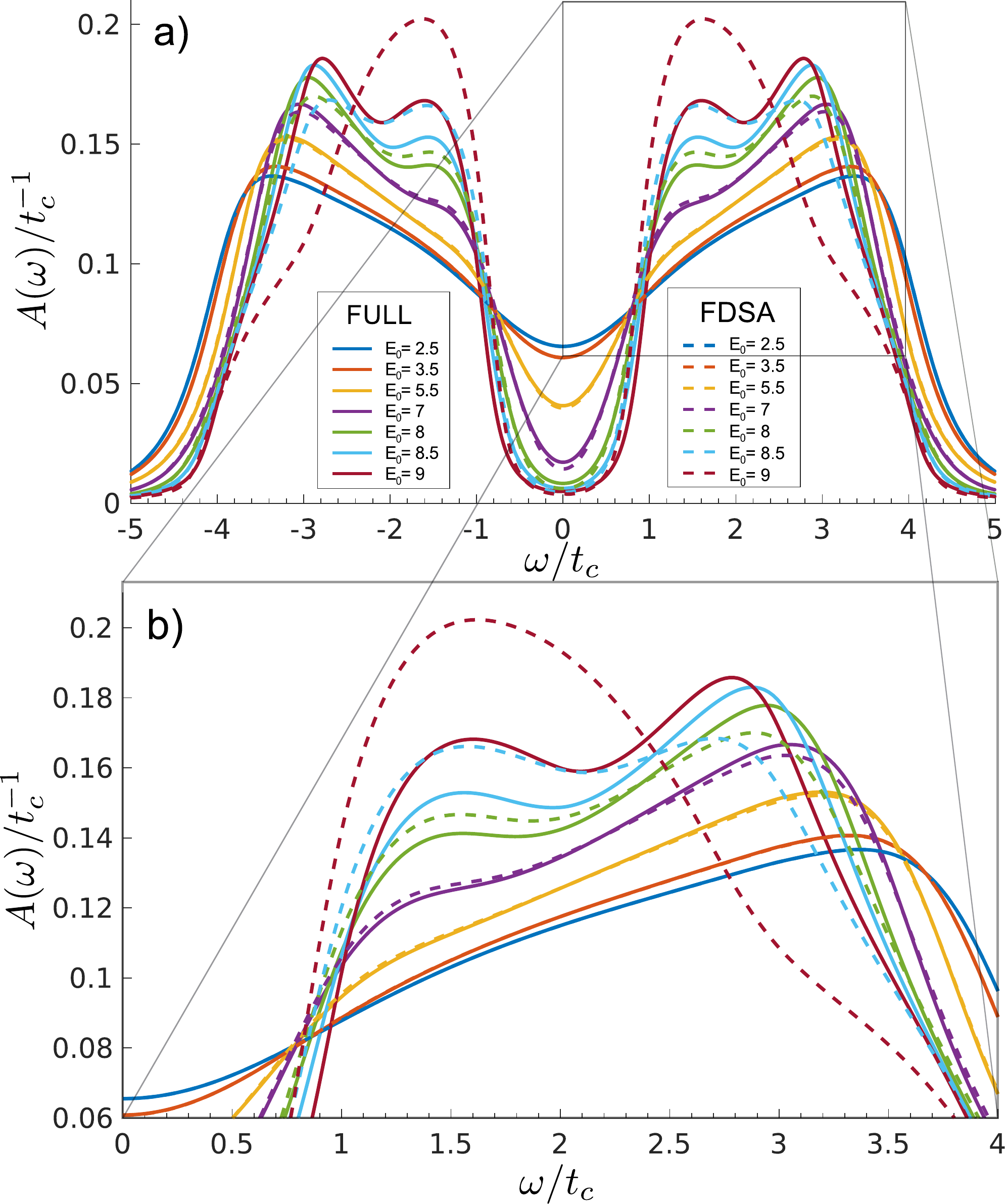}
 \caption{
Test of the FDSA introduced in Sec.(\ref{subsubsec: Time local Floquet DMFT}). a) Time averaged spectral function for different electric field strengths at constant frequency $\Omega = 5$ obtained with IPT. b) The lower plot shows a zoom onto the upper Hubbard band as indicated by the box in the upper plot.
Numerical parameters are the same as in Fig.~\ref{fig:IPT Current}.
}
 \label{fig:IPT DOS}
\end{figure}

\FloatBarrier
For the sake of simplicity and because the nature of the approximation does not depend on it, we present here checks where the total system defined by Eq.~(\ref{eq:Hamiltonian}) is two dimensional. That is, the correlated central region is a Hubbard chain instead of a layer. In Figs.~\ref{fig:IPT Current} and (\ref{fig:IPT DOS}) we plot data for $U=5$, $V_l=V_r=0.1$ and zero temperature $k_{b}T=0$. Fig.~\ref{fig:IPT Current} shows the steady state current density as a function of the effective driving strength $\alpha = \frac{t_cE_0}{\Omega^2}$ for different driving frequencies $\Omega$. In Fig.~\ref{fig:IPT DOS}, we complement this with the results for the local spectral function at $\Omega = 5$ for selected electric field strengths $E_0$. Together, they confirm that the FDSA is justified for the moderate driving intensities and large frequencies,
 which we study within AMEA in this paper.
 The reason why the current approaches zero for the highest two considered $\Omega$'s for large electric fields is due to dynamical localization~\cite{du.ke.86} which localizes the spectrum in frequency (see also Fig.~\ref{fig:IPT DOS}), and therefore suppresses the current for these higher driving frequencies as ``photonic'' excitations are no longer possible within the spectrum. A more detailed analysis and benchmark in different parameter regimes with different impurity solvers is beyond the scope of this work and will be presented elsewhere.

For the sake of completeness, let us briefly recapitulate  the non-equilibrium IPT equations that have been used for this analysis. In short, IPT is second order perturbation theory in the Hubbard interaction $U$. However, in the case of the one band Hubbard model at particle-hole symmetry it turns out to be exact in the limit of infinite interaction as well. It is hence quite reliable in this case, whereas it fails off particle-hole symmetry. In technical terms, IPT is very simple as the self-energy for a given hybridization $\underline{\Delta}(t,t')$ is given by
\begin{equation} \label{eq:IPT self-energy}
 \Sigma^{\stackrel{<}{>}}(t,t') = U^2\mathcal{G}^{\stackrel{<}{>}}_{0}(t,t')\mathcal{G}^{\stackrel{<}{>}}_{0}(t,t')\mathcal{G}^{\stackrel{>}{<}}_{0}(t',t)
\end{equation}
where $G^{</>}$ refer to the lesser/greater GF in the Keldysh formalism and we introduced the so-called Weiss GF
\begin{equation} \label{eq: Weiss GF}
 \underline{\mathcal{G}}^{-1}_{0}(t,t') = \underline{g}_{\text{imp}}^{-1}(t,t') - \underline{\Delta}(t,t').
\end{equation}
\FloatBarrier

\bibliographystyle{/afs/itp.tugraz.at/user/arrigoni/noneq-group/bibtex/prsty} 
%\bibliography{New_References,/afs/itp.tugraz.at/user/arrigoni/noneq-group/bibtex/references_database.bib}
\bibliography{Additional_Refs.bib,references_database.bib}
%\bibliography{New_References,references_database.bib}

\end{document}